\documentclass[conference]{IEEEtran}
\IEEEoverridecommandlockouts
\usepackage{cite}
\usepackage{amsmath,amssymb,amsfonts}
\usepackage{algorithmic}
\usepackage{graphicx}
\usepackage{textcomp}
\usepackage{comment}
\usepackage{adjustbox}
\usepackage{xcolor}
\usepackage{booktabs}
\def\BibTeX{{\rm B\kern-.05em{\sc i\kern-.025em b}\kern-.08em
    T\kern-.1667em\lower.7ex\hbox{E}\kern-.125emX}}
    
\begin{document}

\title{A High-Voltage Characterisation 
Platform For Emerging Resistive Switching Technologies\\

\thanks{ }
}

\author{
    \IEEEauthorblockN{Jiawei Shen$^{*}$,  Andrea Mifsud$^{*\dagger}$, Lijie Xie$^{*}$, Abdulaziz Alshaya$^{*}$, Christos Papavassiliou$^{*}$}
    \IEEEauthorblockA{$^{*}$Department of Electrical and Electronic Engineering, Imperial College London, SW7 2BT, UK\\$^\dagger$Centre for Bio-Inspired Technology, Institute of Biomedical Engineering, Imperial College London, SW7 2AZ, UK\\
    Corresponding author email: jiawei.shen17@imperial.ac.uk
    }%
}%

\maketitle 

\begin{abstract}
Emerging memristor-based array architectures have been effectively employed in non-volatile memories and neuromorphic computing systems due to their density, scalability and capability of storing information. Nonetheless, to demonstrate a practical on-chip memristor-based system, it is essential to have the ability to apply large programming voltage ranges during the characterisation procedures for various memristor technologies. This work presents a 16x16 high voltage memristor characterisation array employing high voltage CMOS circuitry. The proposed system has a maximum programming range of $\pm22V$ to allow on-chip electroforming and I-V sweep. In addition, a Kelvin voltage sensing system is implemented to improve the readout accuracy for low memristance measurements. This work addresses the limitation of conventional CMOS-memristor platforms which can only operate at low voltages, thus limiting the characterisation range and integration options of memristor technologies.

\end{abstract}


\section{Introduction}
The term "memristor", a portmanteau of memory and resistor, was coined by Chua in 1971 \cite{chua}. Starting with a circuit theoretic notion of symmetry and completeness, Chua argued that a two-terminal circuit element whose operation is defined by a relationship between flux linkage and charge  must exist and be a fourth fundamental circuit element. Due to their compact structure and promising compatibility with conventional CMOS manufacturing, memristors and memristor crossbar arrays have been widely utilized in logic circuits\cite{intro_logic_circuit}, neuromorphic computing\cite{intro_nuro}, machine learning \cite{intro_machine} and  image processing applications\cite{intro_image}.

The conventional selector-less memristor crossbar array, which typically consists of orthogonal metal lines termed as word lines and bit lines, offers  low power
consumption and high density memristor integration. However, due to the selector-less configuration, a large sneak current is usually present during the writing and reading operations which  can significantly corrupt the accuracy of cross-point analog resistance measurement \cite{intro_sneak}.
To address this issue, several groups have developed  integrated memristor chips which combine a switch-controlled memristor crossbar array, digital interface circuity and memristor characterisation circuits \cite{onchip_memristor}\cite{onchip_memristor2}. This kind of integrated memristor-CMOS systems offer efficient hardware solutions for different network sizes and applications such as edge computing, and can be scaled up to  larger system sizes.
\begin{figure}[!t]
\centering
\includegraphics[width=0.5\textwidth]{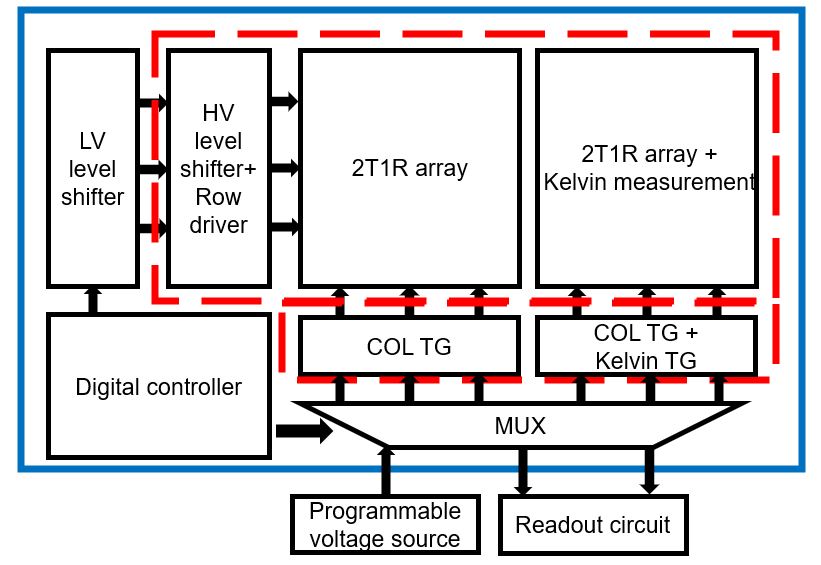}
\caption{\label{fig:chip_level} The top-level architecture of proposed memristor array. The blocks in the red dashed box are the High Voltage circuit and the blue box indicates the entire system implemented on-chip}
\end{figure}

However, these array architectures also have their limitations; (1) their on-chip programming range is  limited by the CMOS supply voltage ($\leq$ 5V in conventional CMOS), so they cannot perform the initial irreversible electroforming process  required for some memristor families to attain their resistive switching property; (2) as-fabricated memristors are perfectly insulating, high amplitude voltage/current pulses across the device are required to initiate a significant  electric conductivity change \cite{electroforming}\cite{electroforming2}\cite{electroforming4}. The necessary amplitude of the electroforming pulse can vary from a few volts to well over 20V\cite{electroforming3}\cite{electroforming4} depending on the structure, size and material of the memristors. It is clear that it is desirable to  integrate all the necessary functions, including the high voltage (HV) electroforming and low voltage readout functions,  on a single chip. By enabling scaling up to arbitrarily large sizes, the fully integrated system introduced here has a great potential to enable the practical implementation of large-scale memristor-based memory. \\

In this paper we report a fully integrated and reprogrammable 2T1R memristor chip designed in the TSMC 180nm BCD Gen 2 technology node. This chip can  electroform devices with voltages up to $|$22V$|$, and can also perform read and write operations in the full voltage range from -22V to 22V. In addition, a Kelvin voltage configuration is also integrated in the 2T1R pixel to enable an accurate measurement of the voltage across the memristor, thus eliminating the error introduced by voltage drops across the 2T selector switches and the column switches. The system designed is highly flexible, supports a wide operating range of programming voltage and is suitable for any type of memristors that can be post-processed on this chip.

\section{CIRCUIT OPERATION}

\begin{figure}[b]
\centering
\includegraphics[width=0.4\textwidth]{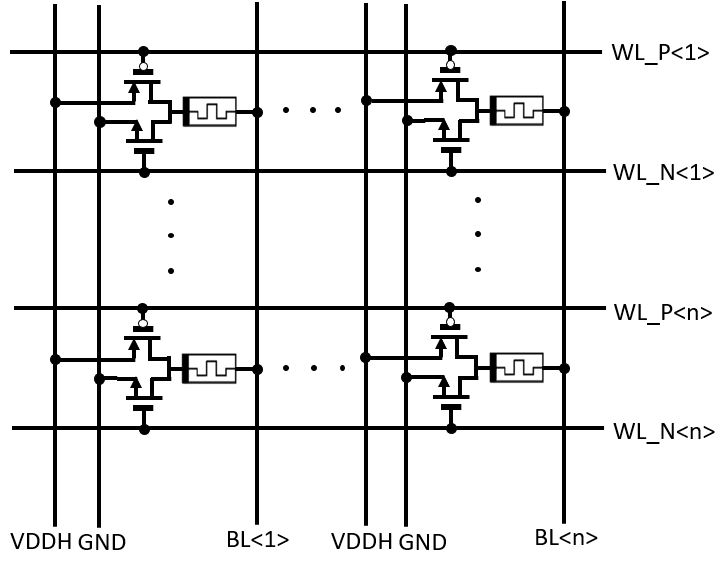}
\caption{\label{fig:pixel} Proposed HV 2T1R array architecture. Horizontal connections for word lines and vertical connections for bit lines}
\end{figure}

\subsection{Chip level architecture}
The block level CMOS chip architecture is shown in Fig. \ref{fig:chip_level}. The  array consists of two types of pixels: 1) HV 2T1R pixel; 2) HV 2T1R pixel in Kelvin configuration, which will be explained in section \ref{section:2t1r} and \ref{Kelvin voltage measurement}.  The array size of each pixel type is 8x16 to form a 16x16 array.

The row circuit includes a HV digital level shifter and a HV digital buffer to drive the HV PMOS  made necessary by the limited $V_{gs}$ swing.  The HV level shifter employs an inverse Schmitt trigger topology to minimize the Miller Plateau effect, which is similar to a previous publication\cite{level_shifter}. The column circuit, on the other hand, uses a HV bi-directional transmission gate to allow current flowing through the Device Under Test (DUT) in both directions.

The main controller activates the relevant rows and columns depending on operational mode and the location of the selected cell in the array. The periphery circuits required to write/read to/from the device are implemented off-chip due to their large size (especially in the HV domain) and also to enable a higher flexibility. All digital signals are generated and decoded by the 1.8V on-chip integrated digital controller.

The inputs to this chip are; (1) digital 1.8V control signals including the address of the DUT, pulse width and operation mode; (2) analog inputs from an off-chip voltage source to establish the desired programming voltage across activated  memristor. The outputs of the proposed system are all analog signals including the current flowing through and the voltage developed across the selected DUT.

\subsection{2T1R Pixel} \label{section:2t1r}
The structure of proposed 2T1R array is illustrated in Fig.\ref{fig:pixel}. The bottom electrode of each memristor is connected to a bit line which will be shared with all the devices of an entire column. Two high voltage transistors are connected to the top electrode of each memristor, and are controlled by separate word lines. The HV NMOS and PMOS are switched separately to control the direction of current flowing through the memristor.  

Because of the typical $V_{GS}$ breakdown constraints of high voltage MOSFETS, the source and bulk terminals of NMOS and PMOS are tied to ground and VDDH respectively so that the voltage of control signals applied to gates are constant and complex $V_{GS}$ protection circuits can be avoided. The gate voltage of the high side PMOS is generated by a high voltage level shifter (VDDH-5V as logic '1' and VDDH as logic '0') to ensure the $V_{GS}\leq5V$ while the gate voltage of NMOS can be simply controlled by a 5V low voltage driver. With an appropriate control sequence and bitline voltages, bidirectional current can flow through the memristor can be established.

\subsection{Kelvin configuration} \label{Kelvin voltage measurement}

\begin{figure}[!b]
\centering
\includegraphics[width=0.5\textwidth]{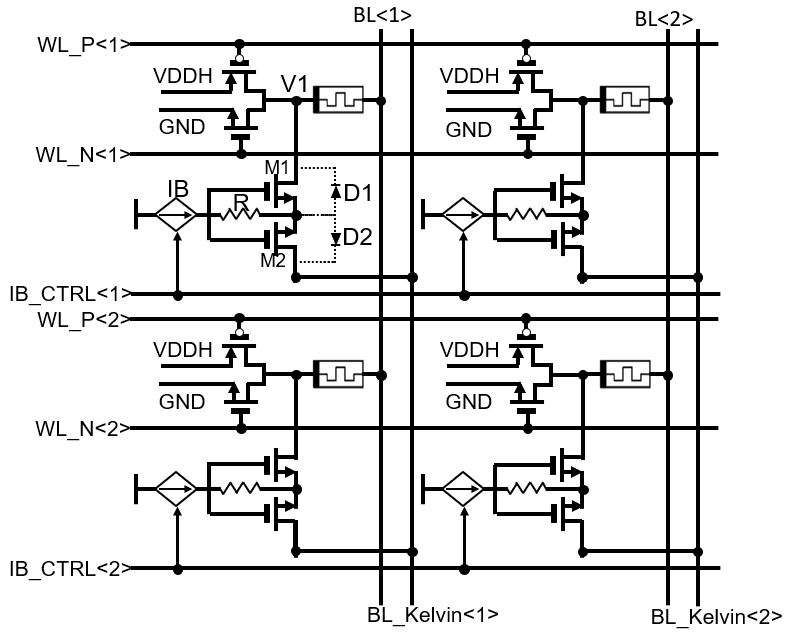}
\caption{\label{fig:kelvin} Proposed 2x2 2T1R pixel with Kelvin voltage sensing nodes}
\end{figure}

\begin{figure}[!t]
\centering
\includegraphics[width=0.4\textwidth]{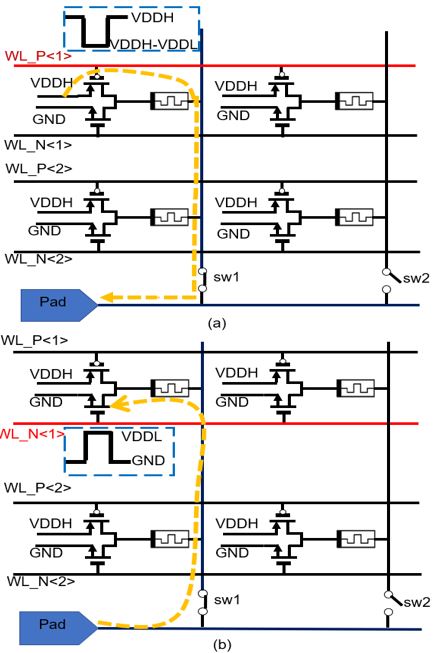}

\caption{\label{fig:programming} The proposed array configuration for the set and reset operations. The activated row is marked in red and applied gate voltage shown in the blue dashed boxes. The direction of current is marked using yellow dashed arrows. (a) reset operation: PMOS activated and current flows through memristor to pad (b) set operation: NMOS activated and current flows through pad to memristor}
\end{figure}

Typical readout circuits measure the total  voltage/current on a memristor including its access network. This way both the selected memristor and on-resistance of activated switches affect the measurement. As the resistance of the memristor decreases, so does  the accuracy of measurements, as the on-resistance of the switches eventually becomes comparable to the LRS (Low Resistance State) of the device. In addition, since a HV supply is used in this chip, the voltage drop across the switches can be significantly higher than the typical low voltage memristor array due to higher currents. To enable a precise measurement of resistance readout, a Kelvin voltage measurement is implemented at the pixel level to accurately sense the voltage across every memristor.
 
Fig. \ref{fig:kelvin} shows the schematic of proposed 2T1R pixel with the voltage sensing node.  M1 and M2 form the transmission gate used  to pass the voltage at top electrode of memristor (V1) to the column-level bitline connection $BL_{kelvin}$. D1 and D2 are the  bulk-drain parasitic diodes of M1 and M2. 
 Current source $I_B$ and R form a gate voltage protection circuit. 
This voltage probe is controlled by a row-level signal $IB_{CTRL}$. When this voltage sensing circuit is activated, a biasing current flows through a passive resistor which generates $V_G=I*R+V_S$. On the other hand, when off, there is no current hence $V_G=V_S$ when not activated. The source  and bulk terminals of M1 and M2 are tied together to ensure symmetry so that $V_{GS}$ and $V_{BS}$ for M1 and M2 will not reach breakdown with the generated gate voltage. 
The voltage across DUT can be accessed by enabling a specific row  $IB_{CTRL}$, $WL_{P,N}$ and connecting a specific column $BL_{Kelvin}$ to pad by decoder.
The current used for controlling Kelvin switches flows through M1 and whichever of the 2T switches is activated, so that it does  not flow through a memristor and affect the readout measurement result. 
For each column, an additional voltage sensing probe with same configuration is employed for BL to measure the voltage at bottom plate of memristor.
The proposed circuit allows simultaneous measurement of $I_{memristor}$,$V_1$ and $V_{BL}$ via three pads.

\subsection{Read/Write operation}

The read/write operation for the chip is performed by selecting the address of the DUT, defining the pulse width and applying the required read/write voltage. 
As shown in Fig. \ref{fig:programming}(a), the reset operation is implemented by pulling the top plate of memristor to VDDH via PMOS and connecting the bottom plate of memristor to the external voltage supply. The voltage across the memristor is $V_{DDH}-V_{PAD}$. The gate voltage of the PMOS swings from $V_{DDH}$ to $V_{DDH}-5v$ with a user-defined pulse width. Similarly, the set operation is implemented by pulling the top plate of the memristor to ground via the NMOS so that the voltage across memristor is $-V_{PAD}$. The gate voltage of the NMOS swings from ground to 5V.
Switch 1 and switch 2 shown in Fig.\ref{fig:programming} have the same configuration as the  Kelvin switch discussed in section \ref{Kelvin voltage measurement} for bi-directional conduction.

\section{RESULTS}
The proposed circuits are designed in a TSMC 180nm BCD Gen 2 technology. The maximum programming voltage for the proposed system is set to 22V. The selections of power supplies satisfy the general requirements for electroforming and achieves optimized compromise between HV device layout size and programming range. 
The simulation is conducted by applying an external voltage supply range from 0-22V to a pad and measuring the current flowing through this pad. The total resistance is calculated from the applied V and measured I values.
\begin{figure}[b]
\centering
\includegraphics[width=0.5\textwidth]{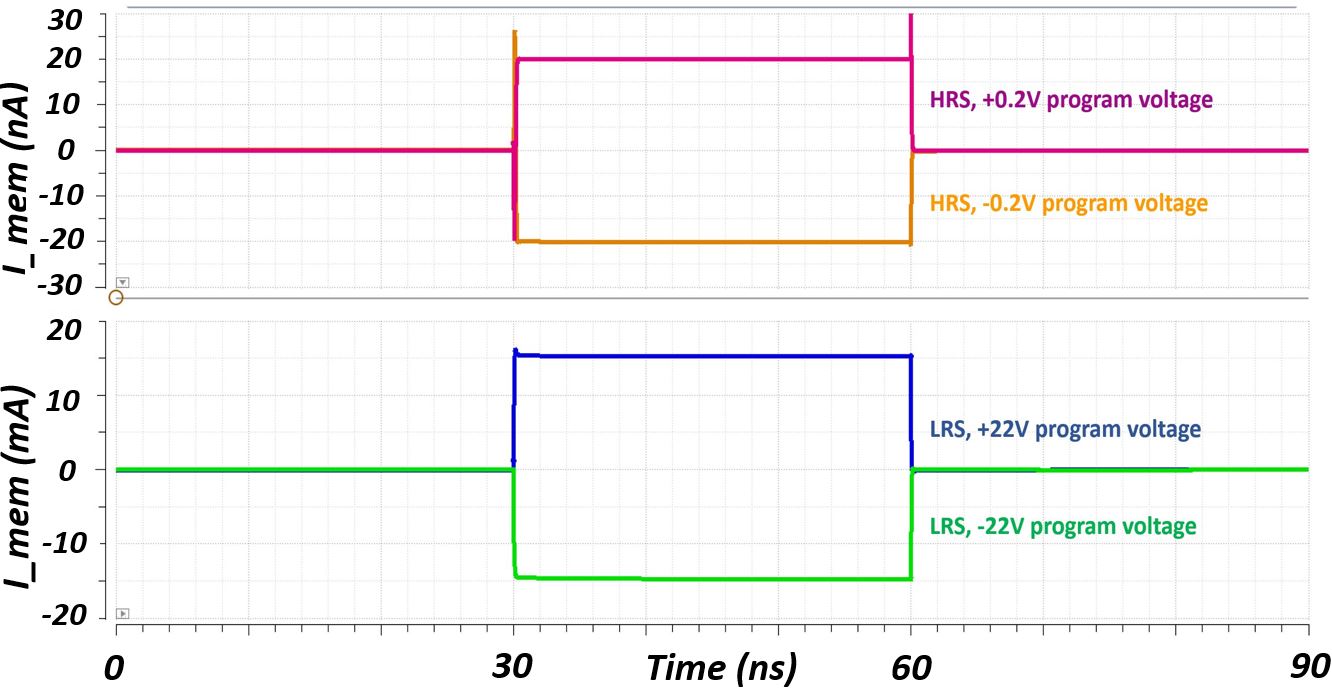}
\caption{\label{fig:transient_current} Current flow through a DUT with 30ns row enable applied (30ns-60ns); (top) Applying a low voltage to a 10M$\Omega$ HRS memristor results into a  small current; (bottom) Applying a high voltage to a 1k$\Omega$ LRS memristor results into a  large current. Y-axis: The current flows through DUT. X-axis: transient time (ns)}
\end{figure}
As this array is designed to characterise memristors and given that there are several types of memristors made of different materials, an ideal resistor was used in simulations instead of a specific behavioral model.

\begin{figure}[t]
\centering
\includegraphics[width=0.5\textwidth]{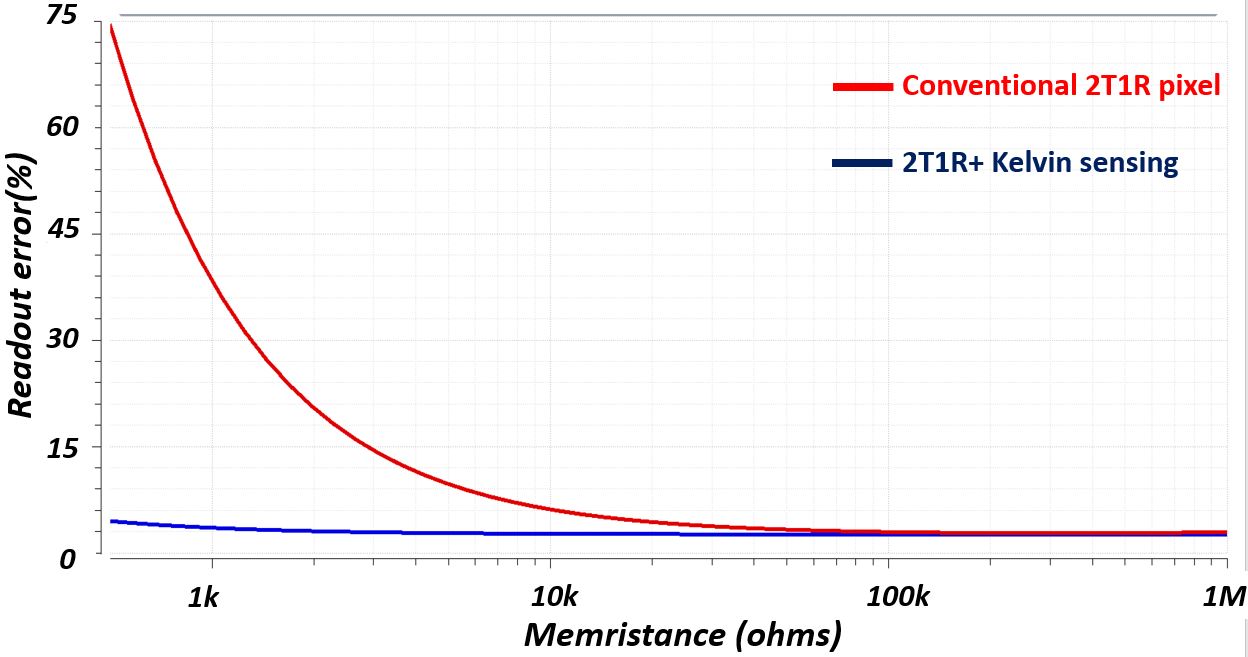}
\caption{\label{fig:readout_error} Comparison of the readout error for a conventional 2T1R cell against a 2T1R cell with Kelvin voltage sensing. Y-axis: The readout error in percentage (\%), X-axis: Sweep of activated memristance (ohms) }
\end{figure}

Fig. \ref{fig:transient_current} shows the transient current flowing through the DUT during a programming phase. The applied pulse width is 30ns which is the minimum achievable pulse width in the designed array. A typical reading voltage of $\pm 0.2V$ is applied to a memristor in $10M\Omega$ HRS resulting in a stable $\pm20nA$ current. To check for the high current situations, the maximum programming voltage of 22V is applied to a memristor in $1K\Omega$ LRS resulting in approximately $\pm15mA$ of current flowing through the device. When the memristors have such low memristance, the on-resistance of the switches is significant. In this example it has caused a large readout deviation from the ideal current of 20mA. Consequently, the accuracy of conventional memristance readout circuits which measure the combination of switch on-resistance and memristance rapidly decreases as the memristor approaches its LRS. 
 \begin{figure}[!b]
\centering
\includegraphics[width=0.5\textwidth]{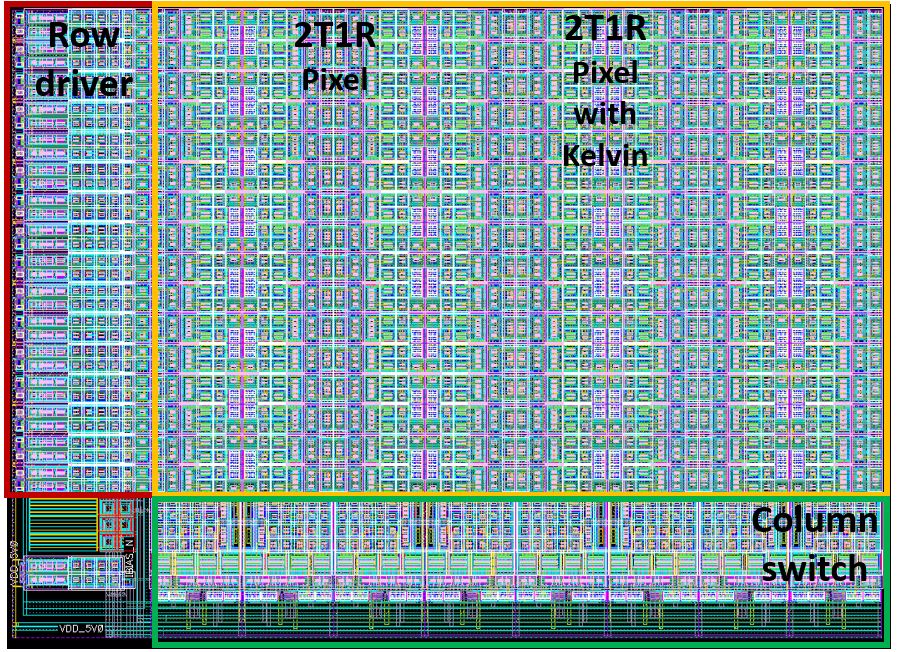}
\caption{\label{fig:chip_layout} The layout of designed chip. Yellow box: pixel array; Red box: Row driver and HV level shifter; Green box: Column switches}
\end{figure}

To address this issue, a Kelvin voltage sensing configuration is utilized to directly measure the voltage drop across a specific memristor. Fig.\ref{fig:readout_error}  depicts  the  improvement  in  accuracy  for  the  2T1R cells  with  Kelvin  voltage  sensing. Note that 
for the conventional 2T1R array, the readout result is derived by the pad voltage, the voltage at the top plate of DUT $V_{TP}$ (0V or 22V) and current measured at readout pad: $R_{mem}=\frac{V_{PAD}-V_{TP}}{I_{PAD}}$.
On the other hand, for the 2T1R cell with Kelvin voltage sensing, the memristance is derived using two voltage measured at two pads $V_{BL\_KELVIN}$, $V_{BL}$ and current measured at  readout pad:$R_{mem}=\frac{V_{BL\_KELVIN}-V_{BL}}{I_{PAD}}$.
For this design, Fig. \ref{fig:readout_error} shows that the 2T1R pixels, have a maximum readout deviation of approximately 75\% at low memristance 500$\Omega$ indicating that most of the total resistance is due to the on-resistance of the switches rather than that of the memristor. By adding Kelvin voltage sensing, the readout deviation is significantly reduced. However it is still higher than 0\%, with it being approximately 3\% throughout the entire memristance range. For HRS, the readout deviation for both systems are approximately same due to the fact that high memristance dominates. The residual error is caused by a small amount of current flowing through M2 in Fig. \ref{fig:kelvin}, which leads to a slight voltage drop. This error can be reduced significantly with direct control of the biasing current.  

The layout of proposed system is shown in Fig.\ref{fig:chip_layout}. The layout size is 1300x940um. The column of 2T1R and 2T1R with Kelvin connection are implemented in an interleaved style to reduce the area of the column circuit as the Kelvin connection requires extra column-level switches. The crucial specifications of proposed array is summarized in Table \ref{table:mm}.

\begin{table}[!t]
\centering
\caption{The specifications of proposed array  }
 \begin{center}
 \begin{tabular}{lcc} 
     \toprule
     Specifications & Unit & Value  \\  
     \midrule
     Programming voltage range & (V) & -22 to 22 \\ 
     \midrule
    2T1R pixel size & ($\mu$m) & 44.5$\times$45 \\
     \midrule
     2T1R pixel size with Kelvin connection & ($\mu$m) &  91.8$\times$46 \\
      \midrule
      Minimum programming pulse width & (ns) & 30 \\
     \midrule
    Maximum tolerable current$^{*}$ & (mA) & 20 \\ 
     \midrule
     Power consumption$^{\dagger}$ & ($\mu$W)& 400 \\
     \midrule
     \multicolumn{3}{l}{ \footnotesize $^{*}$ Max. current through one DUT at room temperature.}\\
    \multicolumn{3}{l}{ \footnotesize $^{\dagger}$ Static power consumption when idle.}\\
    \bottomrule
\end{tabular}	
\label{table:mm}
\end{center}
\end{table}

\section{CONCLUSION}
In this paper, a high voltage memristor-based array based on the 2T1R structure is proposed as part of a characterisation platform for emerging resistive devices. The designed array exhibits a wide programming range ($\pm22V$) and is capable of performing HV functionalities such as on-chip electroforming and I-V sweep. In addition, an innovative pixel structure using a 2T1R cell and Kelvin voltage sensing is proposed to reduce the readout error caused by switches and metal tracks. The two pixel flavours offer a compromise between accuracy and area for large memristor arrays and low resistance measurements. In general, the proposed system is applicable for both volatile and non-volatile resistive memories that require high programming voltage and electroforming.

\end{document}